      \documentstyle[12pt,graphicx]{article}                
\begin{document}
\baselineskip 18pt
\def\today{\ifcase\month\or
 January\or February\or March\or April\or May\or June\or
 July\or August\or September\or October\or November\or December\fi
 \space\number\day, \number\year}
\def\thebibliography#1{\section*{References\markboth
 {References}{References}}\list
 {[\arabic{enumi}]}{\settowidth\labelwidth{[#1]}
 \leftmargin\labelwidth
 \advance\leftmargin\labelsep
 \usecounter{enumi}}
 \def\newblock{\hskip .11em plus .33em minus .07em}
 \sloppy
 \sfcode`\.=1000\relax}
\let\endthebibliography=\endlist
\def\lsim{\ ^<\llap{$_\sim$}\ }
\def\gsim{\ ^>\llap{$_\sim$}\ }
\def\r2{\sqrt 2}
\def\beq{\begin{equation}}
\def\eeq{\end{equation}}
\def\beqn{\begin{eqnarray}}
\def\eeqn{\end{eqnarray}}
\def\rmuu{\gamma^{\mu}}
\def\rmud{\gamma_{\mu}}
\def\PL{{1-\gamma_5\over 2}}
\def\PR{{1+\gamma_5\over 2}}
\def\sinW2{\sin^2\theta_W}
\def\AEM{\alpha_{EM}}
\def\mul{M_{\tilde{u} L}^2}
\def\mur{M_{\tilde{u} R}^2}
\def\mdl{M_{\tilde{d} L}^2}
\def\mdr{M_{\tilde{d} R}^2}
\def\mz2{M_{z}^2}
\def\c2b{\cos 2\beta}
\def\au{A_u}
\def\ad{A_d}
\def\cob{\cot \beta}
\def\v#1{v_#1}
\def\tb{\tan\beta}
\def\epem{$e^+e^-$}
\def\KK{$K^0$-$\bar{K^0}$}
\def\wi{\omega_i}
\def\xj{\chi_j}
\def\Wmu{W_\mu}
\def\Wnu{W_\nu}
\def\m#1{{\tilde m}_#1}
\def\mH{m_H}
\def\mw#1{{\tilde m}_{\omega #1}}
\def\mx#1{{\tilde m}_{\chi^{0}_#1}}
\def\mc#1{{\tilde m}_{\chi^{+}_#1}}
\def\mwi{{\tilde m}_{\omega i}}
\def\mxi{{\tilde m}_{\chi^{0}_i}}
\def\mci{{\tilde m}_{\chi^{+}_i}}
\def\mz{M_z}
\def\sw{\sin\theta_W}
\def\cw{\cos\theta_W}
\def\cb{\cos\beta}
\def\sb{\sin\beta}
\def\rwi{r_{\omega i}}
\def\rxj{r_{\chi j}}
\def\rfp{r_f'}
\def\Kik{K_{ik}}
\def\Fq2{F_{2}(q^2)}
\def\mg{m_{\frac{1}{2}}}
\def\mchi1{m_{\chi}}
\def\tw{\tan\theta_W}
\def\sec2w{sec^2\theta_W}

\begin{center}{\Large \bf  
 Gaugino Mass Nonuniversality and Dark Matter in SUGRA, 
 Strings and D Brane Models} \\
\vskip.25in
{Achille Corsetti\footnote{E-mail: corsetti@neu.edu} and  
Pran Nath\footnote{E-mail: nath@neu.edu}  }

{\it
 Department of Physics, Northeastern University, Boston, MA 02115-5005, USA\\
}

\end{center}

\begin{abstract}  
The effects of nonuniversality of gaugino masses on dark matter 
 are examined within supersymmetric grand unification, and in string 
 and D brane models with R parity invariance. In SU(5) unified models 
 nonuniversality in the gaugino sector can be generated via the gauge 
 kinetic energy function which may depend on the 24, 75 and 200 
 dimensional Higgs representations. We also consider string models  
 which allow for nonuniversality of  gaugino masses and D brane models
 where nonuniversality arises from embeddings of the Standard Model 
 gauge group on five branes and nine branes. It is found that with 
 gaugino mass nonuniversality the range of the LSP mass can be extended 
 much beyond the range allowed in the universal SUGRA case, up to about 
 600 GeV even without coannihilation effects in some regions of the 
 parameter space. 
 The effects of coannihilation are not considered and inclusion of 
 these effects may further increase the allowed neutralino mass range.
 Similarly with the 
 inclusion of gaugino mass nonuniversality, the neutralino-proton 
 ($\chi -p$) cross-section can increase by as much as a factor of 10 
 in some of regions of the parameter space. An analysis of the 
 uncertainties in the quark density content of the nucleon is given and 
 their effects on $\chi -p$  cross-section are discussed. 
 The predictions 
 of our analysis including nonuniversality is compared with the current
  limits from dark matter detectors and implications for future dark 
  matter searches are discussed.  
\end{abstract}

\section{Introduction}
In this paper we study the effects of nonuniversality of gaugino
masses on dark matter in SUGRA models\cite{chams} and
in string and D brane models.
 Nonuniversal gaugino masses arise quite naturally in supergravity
and string unified theories. Thus, 
 in N=1 supergravity the kinetic energy and the mass terms for the
gauge fields and the gauginos are given by\cite{chams} 

\begin{eqnarray}
e^{-1}{\cal L}=&& -\frac{1}{4}\Re\!\left[f_{\alpha\beta}F_{\mu\nu}^{\alpha}
F^{\beta\mu\nu}\right] + \frac{1}{4} i\Im\!\left[
f_{\alpha\beta}F_{\mu\nu}^{\alpha}\tilde{F}^{\beta\mu\nu}\right]
+ \frac{1}{2}\Re\!\left[ f_{\alpha\beta}\left(
- \frac{1}{2}\bar{\lambda}^{\alpha}D\!\!\!\!/\lambda^{\beta}\right) \right]
\nonumber \\
& & - \frac{1}{8} i \Im\!\left[ f_{\alpha\beta} e^{-1}D_{\mu}
(e \bar{\lambda}^{\alpha}\gamma^{\mu}\gamma_5\lambda^{\beta})\right]
+ \frac{1}{4}\bar{e}^{G/2}G^a(G^{-1})^b_a(\partial f^*_{\alpha\beta}
/\partial z^{*b}\lambda^{\alpha}\lambda^{\beta}) + {\rm h.c.}
\end{eqnarray}
Here $\lambda^{\alpha}$ are the gaugino fields, 
 $G = -\ln[\kappa^6 W W^*]- \kappa^2 d$, where
$W$ is the superpotential, $d(z,z^*)$ is the Kahler potential
where $z^a$ are the complex  scalar fields, and  
$\kappa = (8\pi G_N)^{-\frac{1}{2}} = 0.41\times 10^{-18}$ 
GeV$^{-1}$ 
where $G_N$ is Newton's constant.  The gauge kinetic energy function 
$f_{\alpha\beta}$  in general has a non-trivial field dependence
involving fields which transform as a singlet or a non-singlet
irreducible representation of the underlying gauge group. 
 After the spontaneous breaking of the gauge symmetry
at the  unification scale to the Standard Model gauge group 
  $SU(2)_L\times U(1)\times SU(3)_C$, one needs to carry out a rescaling
  of the gauge kinetic energy in the sector of the gauge group that is
  preserved. This rescaling generates a splitting of the 
  $SU(2)_L\times U(1)\times SU(3)_C$ gauge couplings at the 
  unification scale $M_X$ and one has\cite{hill,ref6,das}
\beq
\alpha_i^{-1}(M_X)=\alpha_X^{-1}(M_X) + \sum_r c_{0r} n_i^r
\eeq
where $n_i^r$ characterize the Higgs vacuum structure of the 
irreducible representation r and $c_{0r}$ parametrize its relative 
strength.   
After rescaling the gaugino mass matrix takes the form
\begin{equation}
m_{\alpha\beta}= \frac{1}{4}\bar{e}^{G/2}G^a(G^{-1})^b_a
(\partial f^*_{\alpha\gamma}
/\partial z^{*b})f^{-1}_{\gamma\beta}
\end{equation}
Here one finds that the contribution to nonuniversality of the
gaugino masses is controlled not only by the nature of the GUT sector
but also by the nature of the hidden sector. Because of this the
splitting of the gaugino masses at $M_X$ is characterized by 
a set of parameters $c_r$ different from $c_{0r}$. Thus after the
breaking of the unified gauge group we parametrize the gaugino
masses at $M_X$ by $\tilde m_i(0)$ where\cite{ref6,das}
\beq
\tilde m_i(0)=m_{\frac{1}{2}}(1+ \sum_r c_r n_i^r)
\eeq
The effect of $c_{0r}$ on the gauge coupling unification 
 has been discussed in the previous literature\cite{das} and we do
not discuss it here. In the analysis of this paper we 
assume unification of gauge couplings at $M_X$ and the refinement
of including $c_{0r}$ correction will not have any significant effect
on our analysis.
 
In SU(5) the gauge kinetic energy function $f_{\alpha\beta}$
transforms as the 
symmetric product of ${\bf 24\times 24}$ and contains the following
representations
\beq
({\bf 24}\times {\bf 24})_{symm}={\bf 1}+{\bf 24}+{\bf 75}
+{\bf 200}
\eeq
The singlet leads to universality of the gaugino masses while 
the non-singlet terms will generate nonuniversality. 
We consider models where we have a linear combination 
of the singlet and a non-singlet representation, i.e., 
${\bf 1+24, 1+75}$ or ${\bf 1+200}$. 
The quantities $n_i^r$ for the representations
${\bf 1, 24, 75, 200}$  are listed in 
Table1\cite{ref6,das,anderson1}. 
Some phenomenological aspects  of nonuniversality of gaugino 
masses have recently been discussed\cite{barger,anderson2,huitu}.
We focus here on their effects on event rates in the direct detection
of dark matter (see Ref.\cite{yamaguchi} for previous work on the
effects of gaugino mass nonuniversality on dark matter).  

Techniques for computing 
the event rate in the scattering of neutralinos off nuclear 
targets has been discussed by many authors\cite{jungman}. 
We follow 
here the procedures discussed in Ref.\cite{direct}. 
In our analysis we impose the $b\rightarrow s+\gamma$ 
constraint\cite{bsgamma} and the bounds on SUSY particles from
the Tevatron and LEP\cite{groom}.  
Specifically we take for the lower limits $m_{\chi^+}>94$ GeV,
$m_{\chi}>33$ GeV, $m_{\tilde \tau_R}>71$ GeV, 
$m_{\tilde t}>87$ GeV, $m_{\tilde g}>190$ GeV, $m_h>113.5$ GeV\cite{lep}  and
for the $b\rightarrow s+\gamma$ branching ratio we take a $2\sigma$
 range around the current experiment\cite{alam}, i.e., we take 
 $2\times 10^{-4}<B(b\rightarrow s+\gamma)<4.5 \times 10^{-4}$.
  The quantity that constrains
theory is $\Omega_{\chi} h^2$ where  
$\Omega_{\chi}=\rho_{\chi}/\rho_c$, where $\rho_{\chi}$ is the
neutralino relic density and $\rho_c=3H_0^2/8\pi G_N$ is the critical
matter density, and  h is the value of the Hubble parameter $H_0$
in units of 100 km/sMpc.
The current limit on h from the
Hubble  Space Telescope is $h=0.71\pm 0.03\pm 0.07$\cite{freedman} and
 recent analyses of $\Omega_m$ give\cite{lineweaver} $\Omega_m=0.3\pm 0.08$. 
 Assuming $\Omega_B\simeq 0.05$, one gets 
 \beq
 \Omega_{\chi} h^2=0.126\pm 0.043
 \eeq
 In this analysis we make a  
  a somewhat liberal choice for the error corridor on 
   $\Omega_{\chi} h^2$, i.e., we choose   
   $0.02\leq \Omega_{\chi} h^2\leq 0.3$.
   The choice of a more restricted corridor does not 
   affect the general conclusions arrived at in this analysis.
   In the theoretical computation of the relic density we use
\beqn
\Omega_{\chi} h^2\cong 2.48\times 10^{-11}{\biggl (
{{T_{\chi}}\over {T_{\gamma}}}\biggr )^3} {\biggl ( {T_{\gamma}\over
2.73} \biggr)^3} {N_f^{1/2}\over J ( x_f )}\nonumber\\
J~ (x_f) = \int^{x_f}_0 dx ~ \langle~ \sigma \upsilon~ \rangle ~ (x) GeV^{-2}
\eeqn
where $({{T_{\chi}}\over {T_{\gamma}}}\biggr )^3$ is the
reheating factor,
 $N_f$ is the number of degrees of freedom at the freeze-out 
temperature $T_f$ and  $x_f= kT_f/m_{\tilde{\chi}}$.
In determining  $J(x_f)$ we use the    
    method developed in Ref.\cite{accurate}.
   The role of $J$ in the context of nonuniversalities 
   will be  elucidated in Sec.5. 
 A number of effects on neutralino dark matter have
 already been studied. These include the effects of 
 nonuniversality of the scalar masses at the unification
 scale\cite{nonuni,accomando1}, effects of variations of the
 WIMP velocity\cite{bottino1,roskowski,corsetti}, effects of
 rotation of the galaxy\cite{kk}, 
 effects of CP violation with EDM constraints\cite{cin}, and
 effects of coannihilation\cite{coanni}. 
 The focus of this analysis is on the effects of nonuniversality
 of the gaugino masses on dark matter. In the present analysis
 we do not include the effects of coannihilation. 
 These effects become important when the NLSP mass $m_i$ lies close to
 the LSP mass, i.e., $\Delta_i= (m_i/m_{\chi}-1)<0.1$. 
 We have identified several regions of the parameter space
  where coannihilations involving $\tilde\tau_1$, $\tilde e_R$,
  the light chargino $\chi^+_1$, or the next to the lightest
  neutralino $\chi_2^0$ occur. However, as we stated above 
  we do not consider coannihilation in this paper and thus eliminate
 such regions of the parameter space by imposing the constraint
 $\Delta_i>0.1$. An analysis in this region requires a separate
 treatment and  will be reported elsewhere\cite{corsetti3}.  
 Recent analyses\cite{bottino2,efo} have pointed to the uncertainties 
 in the quark masses and in the quark content of  the nucleon that
 enter in analyses of dark matter. 
 We  give in this paper an independent 
 analysis of the errors in the quark densities and compute their
 effects on dark matter.
 
 The outline of the rest of the paper is as follows: In Sec.2 
 we discuss the basic formulae used to compute the neutralino-proton
 cross-section. An analysis of errors in the quark densities 
 that enter in the scalar $\sigma_{\chi p}$ cross-section is 
 also given. In Sec.3 we first give an analysis of $\sigma_{\chi p}(scalar)$
 for the universal SUGRA case and analyze the effect of errors on the
 quark densities of the proton on it. We then discuss
  three nonuniversal scenarios where we consider  admixtures of
 the singlet with the 24 plet, the 75 plet and the 200 plet 
 representations for the gauge kinetic energy function. 
In Sec.4 we extend the analysis of  $\sigma_{\chi p}(scalar)$ 
to the case  of the O-II string model and  a brane model based on
9 branes and $5_i$ branes. In Sec.5 we discuss the origin of 
the enlargement of the allowed LSP domain consistent with the relic
density constraints due to the presence of nonuniversalities. 
Conclusions are given in Sec.6. In Appendix A we give an analytic
solution of the sparticle masses using the one loop renormalization
group equations including the effect of gaugino mass nonuniversalities.
Using results of Appendix A we compute in  Appendix B the 
effects of nonuniversalities on the $\mu$ parameter.
The analytic results of Appendices A and B provide a deeper 
understanding of the gaugino-mass nonuniversality effects 
discussed in Secs. 3, 4 and 5.

 \section{Neutralino-proton cross-section}
 For heavy target nuclei
such as germanium the neutralino-nucleus scattering cross-section 
is dominated by the scalar part of the neutralino-quark interaction
and it is the quantity $\sigma_{\chi p}(scalar)$ on 
which constraints have been exhibited in the recent 
experimental works\cite{dama,cdms}. For this reason we focus in
this paper on the analysis of $\sigma_{\chi p}(scalar)$.
The basic interaction governing the $\chi -p$ scattering is 
the effective four-fermi interaction given by\cite{chatto}
\beqn
{\cal L}_{eff}=\bar{\chi}\gamma_{\mu} \gamma_5 \chi \bar{q}
\gamma^{\mu} (A P_L +B P_R) q+ C\bar{\chi}\chi  m_q \bar{q} q
+D  \bar{\chi}\gamma_5\chi  m_q \bar{q}\gamma_5 q
+E\bar{\chi}i\gamma_5\chi  m_q \bar{q} q\nonumber\\
+F\bar{\chi}\chi  m_q \bar{q}i\gamma_5 q
\eeqn
where the interaction relevant to our analysis is  
 parametrized by $C$. 
 The $\chi -p$  cross-section 
 arising from scalar interactions is given by 
 \beq
 \sigma_{\chi p}(scalar)=\frac{4\mu_r^2}{\pi}
 (\sum_{i=u,d,s}f_i^pC_i+\frac{2}{27}(1-\sum_{i=u,d,s}f_i^p)
 \sum_{a=c,b,t}C_a)^2
 \eeq
 Here $\mu_r$ is the reduced mass, $f_i^p$ (i=u,d,s quarks)
  are defined by
 \beq
 m_pf_i^p=<p|m_{qi}\bar q_iq_i|p>
 \eeq
 and C is given by
 
 \begin{equation}
  C=C_{h^0}+C_{H^0}+C_{\tilde{f}}
  \end{equation}
   where 
 $C_{h^0},C_{H^0}$ are the contributions from 
  the s-channel $h^0$ and $H^0$ exchanges and 
$C_{\tilde{f}}$ is the contribution from the t-channel sfermion 
exchange. They are given by \cite{chatto}

\beqn
C_{h^0}(u,d)=-(+)\frac{g^2}{4 M_W M^2_{h^0}}
\frac{\cos\alpha(sin\alpha)}{\sin\beta(cos\beta)} Re\sigma
\eeqn

\beqn
C_{H^0}(u,d)=
\frac{g^2}{4 M_W M^2_{H^0}}
\frac{\sin\alpha(cos\alpha)}{\sin\beta(cos\beta)} Re \rho
\eeqn

\beq
C_{\tilde{f}}(u,d)= -\frac{1}{4m_q}\frac{1}
{M^2_{\tilde{q1}}-M^2_{\chi}} Re[C_{qL}C^{*}_{qR}]
-\frac{1}{4m_q}\frac{1}
{M^2_{\tilde{q2}}-M^2_{\chi}} Re[C^{'}_{qL}C^{'*}_{qR}]
\eeq
Here (u,d) refer to the quark flavor, $\alpha$ is the Higgs mixing angle,  
 and $C_{qL}, C_{qL}'$ etc. are as defined in Ref.\cite{chatto}, 
 and  $\sigma$ and $\rho$ are 
defined by

\beq
\sigma= 
 X_{40}^*(X_{20}^*-\tan\theta_W X_{10}^*)\cos\alpha
+X_{30}^*(X_{20}^*-\tan\theta_W X_{10}^*)\sin\alpha
\eeq

\beq
\rho=
- X_{40}^*(X_{20}^*-\tan\theta_W X_{10}^*)\sin\alpha
+X_{30}^*(X_{20}^*-\tan\theta_W X_{10}^*)\cos\alpha
\eeq
 where $X_{n0}$ are the components of the LSP  
 \beq 
\chi=X^*_{10} \tilde B+ X^*_{20}\tilde W_3 + X^*_{30}\tilde H_1
+ X^*_{40} \tilde H_2
\eeq
We discuss now the amount of uncertainty connected with  the 
determination of $f_i^p$.
The quantities that are used as inputs are 
$\sigma_{\pi N}$, x, and $\xi$ defined by

\beq
 <p|2^{-1}(m_u+m_d)(\bar uu+\bar dd|p>=\sigma_{\pi N},
\eeq

 \beq 
 x=\frac{\sigma_0}{\sigma_{\pi N}}=
 \frac{<p|\bar uu+\bar dd-2\bar ss|p>}{<p|\bar uu+\bar  dd|p>}
\eeq
and 
\beq 
 \xi=
 \frac{<p|\bar uu-\bar dd|p>}{<p|\bar uu+\bar  dd|p>}.
\eeq

 We can determine $f_i^p$ in terms of these and find
\beqn 
f_u^p=\frac{m_u}{m_u+m_d}(1+\xi)\frac{\sigma_{\pi N}}{m_p}\nonumber\\
f_d^p=\frac{m_d}{m_u+m_d}(1-\xi)\frac{\sigma_{\pi N}}{m_p}\nonumber\\
f_s^p=\frac{m_s}{m_u+m_d}(1-x)\frac{\sigma_{\pi N}}{m_p}
\eeqn

 A similar analysis holds for the neutralino-neutron scattering
 and one can determine $f_i^n$ in terms of $\xi,x,\sigma_{\pi N}$ as
 follows 
\beqn 
f_u^n=\frac{m_u}{m_u+m_d}(1-\xi)\frac{\sigma_{\pi N}}{m_p}\nonumber\\
f_d^n=\frac{m_d}{m_u+m_d}(1+\xi)\frac{\sigma_{\pi N}}{m_p}\nonumber\\
f_s^n=\frac{m_s}{m_u+m_d}(1-x)\frac{\sigma_{\pi N}}{m_p}
\eeqn 
We note in passing that from Eqs.21  and 22 one has the relation
\beq
f_u^pf_d^p=f_u^nf_d^n
\eeq
which holds independent of the details of the input parameters.
We discuss now the numerical evaluation of $f_i^p$ and $f_i^n$.
The various determinations of $\sigma_{\pi N}$, $\sigma_0$ and
x using analyses of Ref.\cite{gls,gl,borasoy,dong,fukugita,bottino2}
 are summarized in Table2. For $\sigma_{\pi N}$ one finds an
average value of $48\pm 9$ MeV, and for $\sigma_0$ an average
value of $35.5\pm 6$  MeV. These give
 $x=0.74\pm 0.25$. Further, there are two independent lattice gauge
determinations\cite{dong,fukugita} of $y=1-x$ which we list in Table2.
In recording the result of $x$ for Ref.\cite{fukugita} we have reduced
the y value by $35\%$ as 
discussed in Ref.\cite{dong}.
 The average of these
lattice gauge calculations gives $x=0.61\pm 0.08$. Taking the
average yet again of this $x$  and of
$\bar\sigma_0$/$\bar\sigma_{\pi N}$ we get the average 
$\bar x$ listed in Table2. In addition to the above we
need to determine the symmetry breaking parameter 
$\xi$. 
Here as in the work of Ref.\cite{efo} we use the analysis of 
Ref.\cite{cheng} on baryon mass 
splittings to obtain

 \beq
 \xi=\frac{(\Xi^-+\Xi^0-\Sigma^+-\Sigma^-)x}
 {\Xi^-+\Xi^0+\Sigma^++\Sigma^--2m_p-2m_n} 
 \eeq
 where $x$ is as defined by Eq.(19). Numerically one finds
 $\xi=   0.196 x$ and on using Table 2 we find
 \beq
\xi=0.132\pm 0.035
\eeq  
In addition to the above one needs the ratios of the quark masses
for which we use\cite{gl}
\beq
\frac{m_u}{m_d}=0.553\pm 0.043, ~~\frac{m_s}{m_d}=18.9\pm 0.8
\eeq
On using Eqs.(21),(25),(26), and Table 2 we find
\beqn 
f_u^p=0.021\pm 0.004\nonumber\\
f_d^p=0.029\pm 0.006\nonumber\\
f_s^p=0.21\pm 0.12
\eeqn 
Similarly for $f_i^n$ we find
\beqn 
f_u^n=0.016\pm 0.003\nonumber\\
f_d^n=0.037\pm 0.007\nonumber\\
f_s^n=0.21\pm 0.12
\eeqn 
For the more general case of neutralino-Nucleus ($\chi -N$) 
 scattering one has
\beqn
 \sigma_{\chi -N}(scalar)=\frac{4 m_r^2}{\pi}
 (Z\sum_{i=u,d,s}f_i^pC_i+\frac{2}{27}Z(1-\sum_{i=u,d,s}f_i^p)
 \sum_{a=c,b,t}C_a\nonumber\\
 +(A-Z)\sum_{i=u,d,s}f_i^nC_i+\frac{2}{27}(A-Z)(1-\sum_{i=u,d,s}f_i^n)
 \sum_{a=c,b,t}C_a)^2
 \eeqn
Using Eqs.21 and 22 we write 
the above in the form

 \beqn
 \sigma_{\chi-N}(scalar)=\frac{4 m_r^2A}{\pi}
 (\hat f \frac{m_uC_u+m_dC_d}{m_u+m_d}+ 
\hat f \xi \Delta \frac{m_uC_u-m_dC_d}{m_u+m_d} +f C_s\nonumber\\
 +\frac{2}{27}(1-f-\hat f -\hat f\xi \Delta \frac{m_u-m_d}{m_u+m_d})
 \sum_{a=c,b,t}C_a)^2
 \eeqn
where  $\Delta = (2Z-A)/A$, $f=f_s$, $\hat f=\sigma_{\pi N}/m_p$ and
numerically on using Table 2 we have
\beq
\hat f=0.05\pm 0.01
\eeq
We note that while the $\chi -p$ and $\chi -n$ cross-sections  
 depend on $\xi$, the $\xi$ dependent term has a 
 cancellation in  ($\chi -N$) cross-section because of the 
  $(2Z-A)$ factor. Further, if the target 
 nucleus has $A=2Z$, i.e., $\Delta =0$, then
 the $\xi$ dependent term will drop out of the $\chi -N$
 cross-section. Because of the above it is 
 experimentally better to plot the $\chi -N$ cross-section rather
 than the $\chi -p$ cross-section as is currently the
 practice\cite{dama,cdms}. 
 
 \section{Dark matter in GUT models
with gaugino mass nonuniversality}
As discussed at the beginning of this section we consider 
here models where the nonuniversalities arise from admixtures 
of the singlet with the {\bf 24} plet, the {\bf 75} plet and 
the {\bf 200} plet representations. 
 However, we begin first by exhibiting
the result for the universal SUGRA case. The soft SUSY breaking
sector of the theory, under the assumption that  SUSY breaking is 
communicated from the
hidden to the visible sector by gravitational interactions, is 
parametrized in this case by the universal scalar mass $m_0$, 
the universal gaugino mass $m_{\frac{1}{2}}$, the universal trilinear 
coupling $A_0$ all taken at the GUT scale, and $\tan\beta=<H_2>/<H_1>$ 
where $H_2$ gives mass to the up quark and $H_1$ gives mass
to the down quark. Throughout this analysis we assume that
the Higgs mixing parameter $\mu$ (which appears in the superpotential
as $\mu H_1H_2$) is determined via the electro-weak symmetry breaking
constraint. The range of the parameters are limited by a naturalness
constraint. We mean this to imply that $m_0, m_{\tilde g}\leq 1$
TeV, where $m_{\tilde g}$ is the gluino mass, $\tan\beta \leq 25$,
  and $A_0$, or equivalently $A_t$, the value of $A_0$ at the electro-weak
  scale in the top channel, is  limited by the
   electro-weak symmetry breaking constraint.
   For the analysis here we choose 
 $\mu >0$ while for the other $\mu$ sign the allowed 
 parameter space for dark matter is strongly limited
 due to the $b\rightarrow s+\gamma$ constraint\cite{bsgamma}.
 The results for $\mu<0$ look qualitatively different in that
 the cross-sections are significantly smaller.

    In Fig.1 we plot
the maximum and the minimum of $\sigma_{\chi p}(scalar)$ as a function of
$m_{\chi}$ where the parameters are allowed to vary over 
the naturalness range discussed above. The analysis is done for three
sets of $f_i^p$ values corresponding to the corridor given
by Eq.(27). They correspond to
(I)$f_u^p=0.025, f_d^p=0.035, f_s^p=0.33$,
(II)$f_u^p=0.021, f_d^p=0.029, f_s^p=0.21$,
(III)$f_u^p=0.017, f_d^p=0.023, f_s^p=0.09$.
From Fig.1 we see that the different sets can lead to a variation in 
$\sigma_{\chi p}(scalar)$ up to a factor of about 5. 
For the rest of the analysis in this paper  we  use set (II).

Next we consider the $\bf {1+24}$ model.
 We find that in this
case $\sigma_{\chi p}(scalar)$
typically decreases for positive values of
$c_{24}$ and increases with negative values of $c_{24}$.  This  behavior
arises primarily from the dependence of the gaugino-Higgsino components 
$X_{n0}$ of the LSP on the gaugino mass nonuniversality. Thus 
 $X_{n0}$ are sensitive to the gaugino nonuniversality through
their dependence on $\tilde m_1$, $\tilde m_2$ and $\mu$. 
In Fig.2 we exhibit the dependence of $X_{n0}$ on $C_{24}$ for some
typical input values. 
The quantity $\sigma_{\chi p}(scalar)$ depends on the direct product of 
the gaugino and Higgsino components of $\chi$.  Specifically, 
$\sigma_{\chi p}(scalar)$ vanishes if $\chi$ is a pure Bino. 
From Fig.2 we see that 
negative values of $c_{24}$ increase the Higgsino components
and hence increase the neutralino-quark scattering and lead to an
enhancement of $\sigma_{\chi p}(scalar)$
 while the opposite situation is realized 
for positive values of $c_{24}$. This is what is found in Fig.3
where we give a plot of the
 maximum and the minimum of $\sigma_{\chi p}(scalar)$
  for the  cases $c_{24}=-0.08$, $c_{24}=0$ and $c_{24}=0.1$
   when the soft SUSY breaking 
  parameters are varied over the assumed naturalness range as in
  Fig.1. 
  The analysis  shows that for $c_{24}=-0.1$ an enhancement of 
  $\sigma_{\chi p}(scalar)$ by as much as a factor of 5 can occur 
  as  a result of the  gaugino mass nonuniversality and 
  the allowed range of $m_{\chi}$ is also increased in this case
  beyond the range allowed  in the universal SUGRA case.

In Fig.4  we give an analysis of the maximum and the minimum of 
$\sigma_{\chi p}(scalar)$ for the 
 $\bf 1+75$ case for three different values of $c_{75}$, i.e.,
 $c_{75}=-0.06$, $c_{75}=0$ and  $c_{75}=0.04$.
As for the ${\bf 1+24}$ case, $\sigma_{\chi p}(scalar)$ typically
increases for negative values of $c_{\bf 75}$ and decreases for
positive values of $c_{\bf 75}$. Again this can be understood 
by analysing the gaugino-Higgsino components of the LSP as a function 
of $c_{\bf 75}$. Thus here as in the ${\bf 1+24}$  case one finds 
that the Higgsino components
of the LSP increase as $c_{\bf 75}$ decreases and decrease 
as $c_{\bf 75}$ increases. 
Because of this there is an enhancement of $\sigma_{\chi p}(scalar)$ for 
 $c_{75}<0$. A comparison of $c_{75}=0$ and $c_{75}=-0.06$ cases 
 in Fig.4 shows
that an enhancement of $\sigma_{\chi p}(scalar)$ 
up to a factor  of 5 or more occurs in this case. 
 As in the case of 
 ${\bf 1+24}$ here also one finds that the allowed range of
 $m_{\chi}$  consistent with the constraints is 
 extended beyond the values allowed in the universal SUGRA
 case.

 In Fig.5  we give an analysis of the maximum and the minimum of the
  $\sigma_{\chi p}(scalar)$
 for the ${\bf 1+200}$ case when $c_{200}=-0.08$, $c_{200}=0$
 and $c_{200}=0.1$. In this case
the dependence of $\sigma_{\chi p}(scalar)$
 on $c_{200}$ is opposite 
to that one has in  the previous cases, i.e., the ${\bf 1+24}$  
case and the ${\bf 1+75}$ case. Here it is for positive values of 
$c_{200}$ that $\sigma_{\chi p}(scalar)$ increases and it is for 
   negative values of $c_{200}$ that $\sigma_{\chi p}(scalar)$ 
 decreases. The origin of this reversal lies in $n_i^r$
(see Table 1) and can be understood from the dependence of $X_{n0}$
on $c_{200}$. The above implies that the Bino component 
of $\chi$ decreases and the Higgsino components increase for  
$c_{200}>0$ while the opposite situation occurs when $c_{200}<0$.
This dependence implies that $\sigma_{\chi p}(scalar)$
  should increase for $c_{200}>0$  
and decrease for $c_{200}<0$ which is what is observed in Fig.5.
  In this case one finds that $\sigma_{\chi p}(scalar)$
   can increase
  up to a factor of 10 or more because of nonuniversality.
 The analysis also shows that as in the case of ${\bf 1+24}$ and 
${\bf 1+75}$ the allowed range of the LSP mass is
extended beyond what is allowed in the universal SUGRA case.

One can gain a deeper understanding of the dependence of 
$X_{n0}$ on $c_r$ and hence a deeper understanding of the 
dependence of $\sigma_{\chi p}(scalar)$ on $c_r$ from studying the 
dependence of the Higgs mixing parameter $\mu$  on $c_r$.
 To appreciate why $\mu$ is such an important 
parameter in this discussion it is useful to look at 
 large $\mu$, i.e., the case
 $\mu^2>> M_Z^2$. 
 In this limit one finds that the LSP eigenvector is 
 given by
 \beqn
 X_{10}\simeq 1-(M_Z^2/2\mu^2)\sin^2\theta_W,
 ~~X_{20}\simeq -(M_Z^2/2(\tilde m_2-\tilde m_1)\mu)
 \sin 2\theta_W \sin 2\beta,\nonumber\\
 X_{30}\simeq (M_Z/\mu)sin\theta_W sin\beta,
 ~~X_{40}\simeq -(M_Z/\mu)sin\theta_W \cos\beta,
  \eeqn  
  Eq.32 shows that in the large $\mu$ limit $\chi$  is 
  mostly a Bino and the corrections to the pure Bino limit 
  are proportional to $(M_Z^2/\mu^2)$ while the Higgsino components
  are proportional to $(M_Z/\mu)$. As already pointed out the
   $\sigma_{\chi p}(scalar)$ depends on the  
  interference of the gaugino and Higgsino components of the LSP, i.e.,
  $X_{i0}\times X_{\alpha 0}$ (i=1,2; $\alpha=3,4$).
  Clearly then as $|\mu|$ increases we go more deeply into the pure Bino
  region reducing $\sigma_{\chi p}(scalar)$. 
  Likewise as $|\mu|$ decreases $\chi$ develops larger  
  Higgsino components $X_{\alpha 0}$($\alpha=3,4$), even though it
  is still dominantly a Bino, $\sigma_{\chi p}(scalar)$ decreases.
   Thus to gain an insight on the
  effect of $c_r$ on $X_{n0}$ and hence on the effect of $c_r$ on
   $\sigma_{\chi p}(scalar)$  we need to understand how $c_r$ affects $\mu$. 
  We address this topic below.

  In SUGRA models $\mu^2$ is determined via the breaking of
  the electro-weak symmetry and thus depends on the 
   gaugino mass nonuniversality through the Higgs mass parameters
   (see Eqs.39 and 43 in Appendix A and Eq.50 in Appendix B).
   We can understand the effect of nonuniversality on $\mu$ 
   analytically by expanding $\mu$ for the nonuniversal case
   around the universal value using $c_r$ 
   as an expansion parameter 
  \beq
  \tilde \mu^2= \mu^2_{0}+\sum_r \frac{\partial\mu^2}
  {\partial c_r} c_r+ O(c_r^2)
  \eeq   
  where $\mu_0$ is the value of $\mu$ for the universal case.
  Using the analysis of Appendix B 
  the pattern  of breaking in the Higgs structure of ${\bf 24,75,200}$
  shows that 
  \beq
  \frac{\partial \mu^2_{24}}{\partial c_{24}} >0,
  \frac{\partial \mu^2_{75}}{\partial c_{75}} >0,
  \frac{\partial \mu^2_{200}}{\partial c_{200}}<0
  \eeq 
 Thus in the neighborhood of $c_r=0$ a negative
 value of $c_{24}$ gives a smaller value of $|\mu|$ leading
 to larger Higgsino components in $\chi$ and hence a 
 larger $\sigma_{\chi p}(scalar)$
  as is observed in the numerical analysis. A similar situation
 holds for the nonuniversality effects from $c_{75}$. However,
 for the nonuniversality effects from $c_{200}$ an opposite 
 situation holds because of the opposite sign of the derivative term
 as given by Eq.34.  
More generally one  finds the same behavior in a larger $c_r$ domain,
i.e.,  
   $\mu^2_{24}<\mu_0^2$ for $c_{24}<0$, $\mu^2_{75}<\mu_0^2$ for
   $c_{75}<0$, and $\mu^2_{200}<\mu_0^2$ for $c_{200}>0$
   and observations similar to those valid for small $c_r$ also
   apply here.
   These results imply that gaugino nonuniversality which makes 
   $\mu$ small produces a deviation of the LSP from the approximate
   Bino limit in a direction which leads to a larger value of 
   $\sigma_{\chi p}(scalar)$.

  Nonuniversality of the gaugino masses also has implications for
  naturalness. One convenient definition of naturalness is via
  the fine tuning  parameter $\Phi$ introduced in Ref.\cite{chan} 
 defined by  $\Phi =\frac{1}{4}+\frac{\mu^2}{M_Z^2}$. 
 Using this definition one can easily compare values of fine tuning
 for the universal and nonuniversal cases. One finds  
$\Phi_{24}<\Phi_{0}$ ~~$(c_{24}<0)$, ~~
$\Phi_{75}<\Phi_{0} ~~(c_{75}<0)$, 
~~$\Phi_{200}<\Phi_{0}~~(c_{200}>0)$,
where $\Phi_{24}$ is $\Phi$ for the case ${\bf 1+24}$ etc, and 
$\Phi_{0}$ is $\Phi$ for the universal case.
Since the correction to $\mu^2$ is negative for the case
of larger Higgsino components one finds that the deviation from
the approximate Bino limit is in the direction of a smaller value of
$\Phi$ and towards the direction of greater naturalness 
relative to the universal case. Thus a smaller $\mu$ leads to a
 larger $\sigma_{\chi p}$ and a larger detection rate and 
 also makes the model more natural by
making $\Phi$ small. In this sense the more natural the SUSY model the larger
is the detection rate.

\section{Dark matter in string/brane models}
One of the main hurdles
 in the analysis of  SUSY phenomenology based on string models is
 that there is as yet not a full understanding of the breaking of
 supersymmetry here. 
 However, there do exist efficient ways to parametrize SUSY breaking
  and one such parametrization is\cite{brig}

\begin{eqnarray}
F^S=\sqrt{3} m_{\frac{3}{2}} (S+S^*)\sin\theta e^{-i\gamma_S},~~
F^i=\sqrt{3} m_{\frac{3}{2}} (T+T^*)\cos\theta \Theta_i e^{-i\gamma_i}
\end{eqnarray}
where $F^S$ is the dilation VEV, $F^i$ are the moduli VEVs, 
 $\theta$ ($\Theta_i$) parametrizes the Goldstino direction in the
 S ($T_i$) field space and $\gamma_S$ ($\gamma_i$)  is the 
 phase. The $\Theta_i$ obey the constraint 
$\Theta^2_1+\Theta^2_2+\Theta^2_3=1$.
 We begin with an  
example of the O-II string model with the 
   soft SUSY breaking sector parametrized by\cite{brig} 
  
\begin{eqnarray}
\tilde m_i=\sqrt{3} m_{\frac{3}{2}}(\sin\theta e^{-i\alpha_S}
-\gamma_i \epsilon \cos\theta e^{-i\alpha_T})\nonumber\\
 m^2_0=\epsilon' (-\delta_{GS})m^2_{\frac{3}{2}},~~
 A_0=-\sqrt{3}m_{\frac{3}{2}}\sin\theta e^{-i\alpha_S}
 \end{eqnarray} 
Here
$\gamma_1=-\frac{33}{5}+\delta_{GS}, \gamma_2=-1+\delta_{GS},
\gamma_3=3+\delta_{GS}$ 
where  $\delta_{GS}$  is the Greene-Schwarz parameter which is  fixed  
by the constraint of anomaly cancellation in a given orbifold  model. 
 Further, as in the GUT analyses we treat $\mu$ as an  
 independent parameter. The phenomenology of this model has 
 been discussed in Ref.\cite{cdg} and the EDM constraints in
 Refs.\cite{everett,in}. However, in the analysis below we 
 do not impose the accelerator constraints\cite{cdg} and as in the analysis
 for GUT models we do not assume CP violation and
 thus set the CP phases to zero.

The stucture of the soft 
SUSY parameters for the model discussed above shows that the
nonuniversality of the gaugino masses is controlled 
by several
parameters in this case:  $\delta_{GS}$, $\theta$ and $\epsilon$ 
which play a role similar to
the role played by the parameters $c_r$ in the case of the GUT models.
 The presence of several parameters leads to 
 many different possibilities for
 generating gaugino mass nonuniversality. 
In addition, there is a new feature in this
string model, not present in GUT models,  in that the universal scalar 
 $(mass)^2$ at the unification scale, i.e. $m_0^2$, is 
   dependent  on $\delta_{GS}$ which therefore  correlates the
  universal scalar mass with the  gaugino mass  nonuniversality. 
  An analysis of the maximum and 
 the minimum curves for $\sigma_{\chi p}(scalar)$ as a function 
 of $m_{\chi}$ is given in
 Fig.6 when the parameters in the model are varied over
 a range of allowed values. 
 A comparison with Fig.1 shows that 
 $\sigma_{\chi p}(scalar)$ can be larger than 
 for the  universal SUGRA case by a factor of as much as 10.  
 One also finds that the range of the
 neutralino mass extends to about 575 GeV significantly beyond 
 what one finds in the universal SUGRA case even without inclusion 
 of the coannihilation effects.

 We discuss next dark matter for a class of D brane models.
 Over the recent past there has been considerable interest in
the study of  Type IIB orientifolds and their 
compactifications\cite{berkooz,ibanez,brig}.
We consider here models with compactifications on a six
torus of the type $T^6=T^2\times T^2\times T^2$. In models of this type one 
has a set of 9 branes, $7_i$ (i=1,2,3) branes,
$5_i$ branes and 3 branes. This set is further constrained by
the requirement of N=1 supersymmetry which requires that on has 
 either 9 branes and $5_i$ branes, or $7_i$ branes and 3 branes.
 Model building on branes allows an additional flexibility in that
 one can associate different parts of the Standard Model gauge 
 group with different branes. In Ref.\cite{everett,accomando} a 
 brane model using two five branes $5_1$ and $5_2$ was investigated 
 while in  Ref.\cite{in} models using 
 9 brane and $5_i$ brane were investigated. We  pursue  here the 
 implications of the  latter possibility\cite{in}. 
 In one of the models of Ref.\cite{in}   
 the Standard Model gauge group is associated with   
 the branes in the following way:  the $SU(3)_C\times U(1)_Y$ 
 is associated with the 9 brane
 while $SU(2)_L$ is associated with the $5_1$ brane. Further, it is
 assumed that the $SU(2)_R$ singlets are associated with the 
 9 brane while the $SU(2)_L$ doublets are associated  with the
 intersection of 9 brane and $5_1$ brane.  
 The soft SUSY breaking sector of this $9-5_1$ brane model is then
 given as follows: The 
 gaugino masses $\tilde m_i$ (i=1,2,3)
 corresponding to the gauge group $SU(3)$, $SU(2)$ and $U(1)$
 are parametrized by\cite{in} 
\begin{eqnarray}
\tilde {m}_1=\sqrt{3} m_\frac{3}{2}\sin\theta e^{-i\gamma_S}
=\tilde m_3=-A_0,~~
\tilde {m}_{2}=\sqrt{3} m_\frac{3}{2}\cos\theta 
\Theta_1 e^{-i\gamma_1}
\end{eqnarray}
while the $SU(2)_L$ singlets  are parametrized by $m_9$ and
$SU(2)_L$ doublets are parametrized by $m_{95_1}$ where\cite{in}
\begin{equation}
m_9^2=m^2_{\frac{3}{2}}(1-3\cos^2\theta \Theta_1^2),~~
m_{95_1}^2=m^2_{\frac{3}{2}}(1-\frac{3}{2}\cos^2\theta (1-
\Theta^2_1)).
\end{equation}
Here the $\theta$ and $\Theta_1$ are the directions of the 
Goldstino in the dilaton and the moduli VEV space as discussed earlier.
To avoid generating tachyons in the theory one needs to impose
the constraint $cos^2\theta\Theta_1^2<1$. 
A second version of this model was also discussed in Ref.\cite{in}.
 Here one associates
 $SU(3)\times U(1)_Y$ with the $5_1$ brane
 while the $SU(2)_L$ with the 9 brane, and  
 assumes that the $SU(2)_R$ singlets were associated with the 
 $5_1$  brane while the $SU(2)_L$ doublets are associated  with the
 intersection of 9 brane and $5_1$ brane as before.
 The soft SUSY breaking sector of this model can be gotten
 from the model discussed above by the interchange
 $\cos\theta \Theta_1\leftarrow\rightarrow \sin\theta$. 
 In the analysis of this paper we  focus on the version
 of the model given by Eqs.37 and 38. As usual we assume that the
parameter $\mu$ is free and we determine it via
the constraint of the radiative breaking of the electro-weak
symmetry.  Again as in other cases we have considered we set
the CP phases to zero.
  Eqs.37 and 38 show that one has  
 nonuniversality at the unification scale both in the scalar
 sector  as well as in the gaugino sector. The limit of universal 
 scalar mass corresponds to $\Theta_1=1/\sqrt 3$.
  Our focus in this paper is on the nonuniversality 
 in the gaugino sector, and so for the numerical analysis we set 
 $\Theta_1=1/\sqrt 3$. 
 In this case the numerical analysis shows that the 
 allowed neutralino mass range
  extends up to about 650 GeV.

  \section{Gaugino mass nonuniversality and LSP mass range} 	
  In SUGRA models with universal boundary conditions at the
  GUT scale, the allowed range of the LSP typically does
  not exceed 200 GeV and with imposition of additional  
  constraints it is  often significantly less. As discussed in 
  Secs.3 and 4 in the 
  presence of gaugino mass nonuniversality one finds 
  that the allowed LSP range  is increased.
   For the ${\bf 1+24}$ case
  the allowed LSP range extends to about $220$ GeV for the case
  $c_{24}=-0.1$. For the ${\bf 1+75}$ case one finds that 
$c_{75}=-0.06$ gives an LSP range which extends to 250 GeV,
while for the ${\bf 1+200}$ case with $c_{200}=0.1$ the allowed
LSP range extends to 375 GeV. A similar situation holds for
the string/brane models. Here one finds that the allowed LSP range 
can extend up to 600 GeV. These extended regions arise even without
inclusion of coannihilation effects which is known to extend the allowed
regions also up to about 700 GeV\cite{coanni}.  
In the regions of the parameters space
considered the effects of coannihilations would infact be negligible since
we are considering only those configurations for which
$\Delta_i>0.1$. The specific mechanism by which the neutralino mass range
is extended is also different than in the case of coannihilation.
Thus for the case of coannihilation the increase in the allowed LSP range
occurs as a consequence of a coupled channel effect while in the
the case of nonuniversalities
the extension of the allowed region of the LSP 
arises due to a significant increase in the value  of J 
for certain ranges of nonuniversalities. Thus one may expand J
as the sum over the final state channels in the $\chi\chi$
annihilation so that $J$=$J(\tilde f\tilde f)+J(WW)+J(ZZ)$+
$J(Zh^0)$+$J(ZH^0)+etc$.
The region of the nonuniversality parameter space 
which leads to an enhancement of $\sigma_{\chi p}$ 
  can also lead to an enhancement of 
  the cross-sections for the $\chi\chi$ annihilation into the final 
  states $W^+W^-,ZZ, Zh, ZH$ etc and hence to an increase in J
  which leads to a decrease in the relic density down to 
permissible limits consistent with constraints.
Thus regions of the parameter space which would otherwise be 
excluded are now included when the gaugino mass nonuniversalities 
are included. 

\section{Conclusion}
In this paper we have analyzed the effects of the nonuniversality
of the gaugino masses on neutralino dark matter in
SUGRA, string and D brane models under the constraint
of R parity invariance. 
It is found that nonuniversality effects can enhance the 
$\chi -p$ cross-section for scalar interactions 
by as much as a factor of 10.
We also carried out an analysis of the uncertainties in 
the numerical determination of 
  $f_i^p$ (i=u,d,s) and find that with the current state
of uncertainties the $\chi -p$ cross-section cannot be pinned down to
better than a factor of about 5.
 Our analysis of the gaugino mass  
nonuniversality also exhibits another important phenomena, i.e.,
that the allowed range of the neutralino mass can be extended up 
to about 600 GeV even without inclusion of the coannihilation effects.
 The effects of coannihilation were not considered 
and inclusion of 
 these effects may further increase the allowed neutralino mass range.
  The extended LSP mass range  should be
of interest in the experimental searches for dark matter in 
the current dark matter detectors\cite{dama,cdms,spooner}
 and in the design of new dark matter 
detectors which are at the planning stage\cite{baudis}.

 \noindent
 {\bf Acknowledgements}\\ 	   	
  This research was supported in part by NSF grant 
PHY-9901057.\\
 
\noindent
{\bf Appendix A: Effects of gaugino mass nonuniversality on
 sparticle masses}\\
  In this appendix we give  analytic solutions to the one
  loop renormalization group equations including the effect 
  of gaugino mass nonuniversality. The analytic 
  formulae for these with universal boundary conditions were
  given in Ref.\cite{ilm}  and for nonuniversality in the 
  scalar mass sector in Ref.\cite{nonuni}. Here we limit
  ourselves to the gaugino mass 
  nonuniversality. These formulae are found useful
  in gaining an analytic understanding of the nonuniversality
  effects.  The one loop RG formulae are given in several papers
  (see, e.g. Ref.\cite{ilm}) and we do not reproduce them here. 
  Rather we discuss the solutions
  under the boundary conditions where $m_0$ is the
  universal scalar mass, $\tilde m_i(0)$ (i=1,2,3) 
  are the gaugino masses for the gauge group sectors $U(1),
  SU(2), SU(3)$
   and $A_0$ is the universal trilinear 
  coupling all taken at the unification scale. The Higgs mass
  parameters, the trilinear couplings, and  
  the squark and slepton masses at the electro-weak scale are 
  all sensitive to the effect of gaugino mass nonuniversality.
   The simplest case is that  of the mass parameter $m_{H_1}^2$ for
   the $H_1$ Higgs which couples to the down quark. Here one finds
  
\beq
m_{H_1}^2=m_0^2+\tilde\alpha_G (\frac{3}{2} {\tilde f}_2(t) +
\frac{3}{10} {\tilde f}_1(t))m_{\frac{1}{2}}^2
\eeq
where   
  
\beq
 {\tilde f}_i(t)=\frac{1}{\beta_i}(1-\frac{1}{(1+\beta_it)^2})
 (\frac{\tilde \alpha_i(0)}{\tilde \alpha_G})
 (\frac{\tilde m_i(0)}{m_{\frac{1}{2}}})^2
\eeq
Here $t=ln(M_X^2/Q^2)$, 
$\beta_i=(b_i/4\pi)\tilde \alpha_i(0)$ where $b_i=(33/5,1,-3)$
for $U(1)$, $SU(2)$, and $SU(3)$, and $\tilde \alpha_i(0)=\alpha_i/4\pi$.  
The ${\tilde f}_i(t)$ contain the nonuniversality effects.
The evolution of the mass parameter for the Higgs   
$H_2$ involves the evolution of the trilinear coupling 
 in the stop channel at the electro-weak scale 
 and this coupling is given by 

\beq
A_t(t)=\frac{A_0}{1+6Y_0F}+\mg (\tilde H_2-\frac{6Y_0\tilde H_3}
{1+6Y_0F})
\eeq
where
\beqn
\tilde H_2=\tilde \alpha_G(\frac{16}{3}\tilde h_3+3\tilde h_2+
\frac{13}{15} \tilde h_1),~~
\tilde H_3=\int_0^{t} E(t)\tilde H_2(t)\nonumber\\
\tilde h_i=\frac{t}{1+\beta_i t}(\frac{\tilde \alpha_i(0)}{\tilde \alpha_G})
 (\frac{\tilde m_i(0)}{m_{\frac{1}{2}}})
\eeqn
 where $Y_0$ is the top Yukawa coupling at the GUT scale and 
 the functions E and F are as defined in Ref.\cite{ilm}. We note 
 that the first term on the right hand side of Eq.(40) which arises 
 purely from the top Yukawa coupling evolution is unaffected by
   nonuniversality while the second term in affected through the
   modification of $\tilde h_i$.  For the  mass parameter $m_{H_2}$ 
  for Higgs  $H_2$ that couples with the top one finds 
 
\beq
  m_{H_2}^2=\mg^2{\tilde e}(t)
+A_0m_0\mg {\tilde f}(t)+m_0^2(h(t)-k(t)A_0^2)
\eeq	 
where the functions  $h(t)$ and $k(t)$ are 
unaffected by nonuniversality and are as given in Ref.\cite{ilm}
while the functions $\tilde e$ and $\tilde f$ are modified due to
nonuniversality. $\tilde e$ and $\tilde f$ are given by 
\beq
{\tilde e}(t)=\frac{3}{2}[\frac{\tilde G_1+Y_0 \tilde G_2}{D(t)}
+\frac{(\tilde H_2+6Y_0\tilde H_4)^2}{3D(t)^2}+\tilde H_8]
\eeq
and
\beq
{\tilde f}(t)=-\frac{6Y_0\tilde H_3(t)}{D(t)^2};~~
D(t)=(1+6Y_0 F(t))
\eeq
Here the various tilde functions containing the 
nonuniversality  are defined below 	
\beqn
\tilde G_1(t)={\tilde F}_2(t)-\frac{1}{3}(\tilde H_2)^2\nonumber\\
\tilde G_2(t)=6{\tilde F}_3(t)-\tilde F_4(t) -4 \tilde H_2(t)
\tilde H_4(t)
+2 F (\tilde H_2)^2 -2 \tilde H_6(t)\nonumber\\
\tilde F_3(t)=F(t){\tilde F}_2(t)-
\int_0^t dt'E(t'){\tilde F}_2(t'),~~
 \tilde F_4(t)=\int_0^tdt' E(t')\tilde H_5(t')\nonumber\\
\tilde H_4(t)=F(t)\tilde H_2(t)-\tilde H_3(t),~~
\tilde H_5(t)=\tilde \alpha_G(-\frac{16}{3}{\tilde f}_3(t)+ 6{\tilde f}_2(t)
-\frac{22}{15}{\tilde f}_1)\nonumber\\
\tilde H_6(t)=\int_o^t dt'(\tilde H_2)^2 E(t'),~~
\tilde H_8(t)=\tilde \alpha_G(-\frac{8}{3}{\tilde f}_3(t)+ {\tilde f}_2(t)
-\frac{1}{3}{\tilde f}_1)
\eeqn	
The squark and slepton masses are also affected by nonuniversality.
For the up squarks in the first two generations one finds
\beqn
m^2_{\tilde u_{iL}}(t)=m_0^2+m_{ui}^2
+\tilde\alpha_G [\frac{8}{3}{\tilde f}_3
+\frac{3}{2}{\tilde f}_2+ 
\frac{1}{30}{\tilde f}_1]m_{\frac{1}{2}}^2
+(\frac{1}{2}-\frac{2}{3}\sin^2\theta_W) M_Z^2 \cos 2\beta\nonumber\\
m^2_{\tilde u_{iR}}(t)=m_0^2+m_{ui}^2
+\tilde\alpha_G [\frac{8}{3}{\tilde f}_3+\frac{8}{15}{\tilde f}_1]m_{\frac{1}{2}}^2
+\frac{2}{3}\sin^2\theta_W M_Z^2 \cos 2\beta
\eeqn
The analysis  of the first two generations of the down quarks
and of the sleptons is similar. 
Finally we discuss the third generation squarks. Here one finds 
\beqn
m^2_{\tilde b_{L}}(t)=\frac{1}{2}m_0^2+m_{b}^2+\frac{1}{2}m_U^2+
+\tilde\alpha_G [\frac{4}{3}{\tilde f}_3+ 
\frac{1}{15}{\tilde f}_1]m_{\frac{1}{2}}^2
+(-\frac{1}{2}+\frac{1}{3}\sin^2\theta_W) M_Z^2 \cos 2\beta\nonumber\\
m^2_{\tilde b_{R}}(t)=m_0^2+m_{b}^2
+\tilde\alpha_G [\frac{8}{3}{\tilde f}_3+\frac{2}{15}{\tilde f}_1]m_{\frac{1}{2}}^2
-\frac{1}{3}\sin^2\theta_W M_Z^2 \cos 2\beta\nonumber\\
m^2_{\tilde t_{L}}(t)=m_Q^2+m_{t}^2
+(\frac{1}{2}-\frac{2}{3}\sin^2\theta_W) M_Z^2 \cos 2\beta\nonumber\\
m^2_{\tilde t_{R}}(t)=m_U^2+m_{t}^2
+\frac{2}{3}\sin^2\theta_W M_Z^2 \cos 2\beta
\eeqn
where $m_U^2$ and $m_Q^2$ are defined by 
\beqn
m_U^2=\frac{1}{3} m_0^2+\frac{2}{3} {\tilde f} A_0m_{\frac{1}{2}}
-\frac{2}{3}kA_0^2+ \frac{2}{3}hm_0^2+
[\frac{2}{3}\tilde e+\tilde\alpha_G(\frac{8}{3}{\tilde f}_3-{\tilde f}_2
+\frac{1}{3}{\tilde f}_1)]\nonumber\\
m_Q^2=\frac{2}{3} m_0^2+\frac{1}{3} {\tilde f} A_0m_{\frac{1}{2}}
-\frac{1}{3}kA_0^2+ \frac{1}{3}hm_0^2+
[\frac{1}{3}\tilde e+\tilde\alpha_G(\frac{8}{3}{\tilde f}_3+{\tilde f}_2
-\frac{1}{15}{\tilde f}_1)]
\eeqn
All the sparticle mass relations limit to the universal 
case\cite{ilm} when we set\\
 $\alpha_i(0)\tilde m_i(0)/\alpha_G m_{\frac{1}{2}}=1$. 

\noindent
{\bf Appendix B : nonuniversality effects on $\mu$}\\
Since $\mu$ is determined via the constraint of electro-weak 
symmetry breaking it is sensitive to the gaugino mass nonuniversality.
Thus one has
\beq 
\mu^2=(m_{H_1}^2-m_{H_2}^2\tan\beta^2)(\tan\beta^2-1)^{-1}
-\frac{1}{2}M_Z^2+ \Delta \mu^2
\eeq
where  $\Delta\mu^2$ is the loop correction. From the above  one
finds 
\beq
   \frac{\partial \mu^2}{\partial c_r}
   =(t^2-1)^{-1}(\mg^2g_r'-t^2(\mg^2e_r'+A_0m_0\mg f_r'))
   + \frac{\partial\Delta \mu^2}{\partial c_r}
  \eeq   
   Here 
$ g_r'=\frac{\partial \tilde g}{\partial c_r}$ where  
$\tilde g=\tilde \alpha_G(\frac{3}{2}\tilde f_2+\frac{3}{10}\tilde f_1)$,
and $f_r'=\frac{\partial \tilde f}{\partial c_r}$.
 For large $\tan\beta$ Eq.(51) reduces down to 
\beq
   \frac{\partial  \mu^2}{\partial c_r}
   =-(\mg^2e_r'+A_0m_0\mg f_r')
   + \frac{\partial\Delta \mu^2}{\partial c_r}
  \eeq   
 These results lead to Eq.(34). \\

\noindent
{\bf Tables:}\\

\begin{center} 

\begin{tabular}{|c|c|c|c|}
\multicolumn{4}{c}{Table~1: Nonuniversalities at $M_X$.  } \\
\hline
  SU(5) rep & $n_1^r$ & $n_2^r$& $n_3^r$ \\
\hline
 {\bf 1} & 1 &1&1 \\
 \hline
 {\bf 24} & -1 &$-3$&$2$\\
 \hline
 {\bf 75} &-5&3 &$1$\\
 \hline
 {\bf 200} & 10 &2& 1\\
 \hline
\end{tabular}\\
\noindent

\end{center}

\begin{center} 

\begin{tabular}{|c|c|c|c|c|c|}
\multicolumn{6}{c}{Table~2: Uncertainties in $\sigma_{\pi N}$,
$\sigma_0$ and x  } \\
\hline
  Ref. & $\sigma_{\pi,N}$ (MeV) & Ref. &$\sigma_0$ (MeV)& Ref. & x\\
\hline
 \cite{gls} & $45\pm 8$  & \cite{gl} & $35\pm 5$&
 { $\bar\sigma_0$}/{$\bar\sigma_{\pi N}$}  & $0.74\pm 0.25$\\
 \hline
\cite{borasoy} & $48\pm 10$ & \cite{borasoy}& $36\pm 7$&\cite{dong}  & $0.64\pm 0.03$ \\
 \hline
 \cite{fukugita}  & $50\pm 10$  &{$\bar\sigma_0$} & $35.5\pm 6$&\cite{fukugita}
  & $0.57\pm 0.1$\\
 \hline
 \cite{bottino2} & $49\pm 8$ & & &$\bar x$ & $0.67\pm 0.18$\\
 \hline
  {$\bar\sigma_{\pi N}$} & $48\pm 9$ &  &  & &  \\
 \hline
\end{tabular}\\
\noindent

\end{center}

\noindent
{\bf Figure captions}\\
Fig.1: Exhibition of the dependence of $\sigma_{\chi p}(scalar)$ 
on the uncertainties in $f_u^p$, $f_d^p$ and $f_s^p$ for the 
universal case. The plots 
are for three sets of $f_i^p$ discussed in Sec.3. The parameter
space spanned is as discussed in the text.\\

\noindent
Fig.2: Plot of gaugino-Higgsino components $X_{n0}$ (n=1,2,3,4) 
as a function of $c_{24}$ for the
input data  $m_0=51$ GeV, $\mchi1$ =70 GeV,  $\tan\beta=10$,
$A_t/m_0=-7$  where  $X_{20}=-|X_{20}|$ and 
$X_{40}=-|X_{40}|.$\\

\noindent
Fig.3: Plot of the maximum and the minimum curves  
$\sigma_{\chi p}(scalar)$
as a function of $m_{\chi}$ 
for the case when one considers admixtures of ${\bf 1+24}$
representations with $c_{24}=-0.1$ (dashed), $c_{24}=0$ (solid)
and  $c_{24}=0.08$ (dotted) when the other parameters 
are varied over the assumed naturalness range as discussed in the 
text. \\

\noindent
Fig.4: Plot of the maximum and the minimum curves  
$\sigma_{\chi p}(scalar)$
as a function of $m_{\chi}$ 
for the case when one considers admixtures of ${\bf 1+75}$
representations with $c_{75}=-0.06$ (dashed), $c_{75}=0$ (solid)
and  $c_{75}=0.04$ (dotted) when the other parameters 
are varied over the assumed naturalness range as discussed in the 
text. \\

\noindent
Fig.5: Plot of the maximum and the minimum curves  
$\sigma_{\chi p}(scalar)$
as a function of $m_{\chi}$ 
for the case when one considers admixtures of ${\bf 1+200}$
representations with $c_{200}=0.1$ (dashed), $c_{200}=0$ (solid)
and  $c_{200}=-0.08$ (dotted) when the other parameters 
are varied over the assumed naturalness range as discussed in the 
text. \\

\noindent
Fig.6: Plot of the maximum and the minimum curves for  
$\sigma_{\chi p}(scalar)$
 as a function of $m_{\chi}$ 
for the heterotic string model O-II. The range of parameters 
consists of $\epsilon,\epsilon'$ in the range 0.0025-0.01,
$\theta$ in the range 0.1-1.6, $\delta_{GS}$ in the range
-1 to -10, $m_{3/2}$ in the range up to 2 TeV, and
values of $\tan\beta$  range up to 25. \\

\noindent
Fig.7: Plot of the maximum and the minimum curves for 
$\sigma_{\chi p}(scalar)$
 as a function of $m_{\chi}$ 
for the $9-5_1$ D brane model. The range of parameters consists of
$\theta$ in the range $0.1-1.6$, $m_{3/2}$ in the range up to
2 TeV, and values of $\tan\beta$ range up to 25. \\

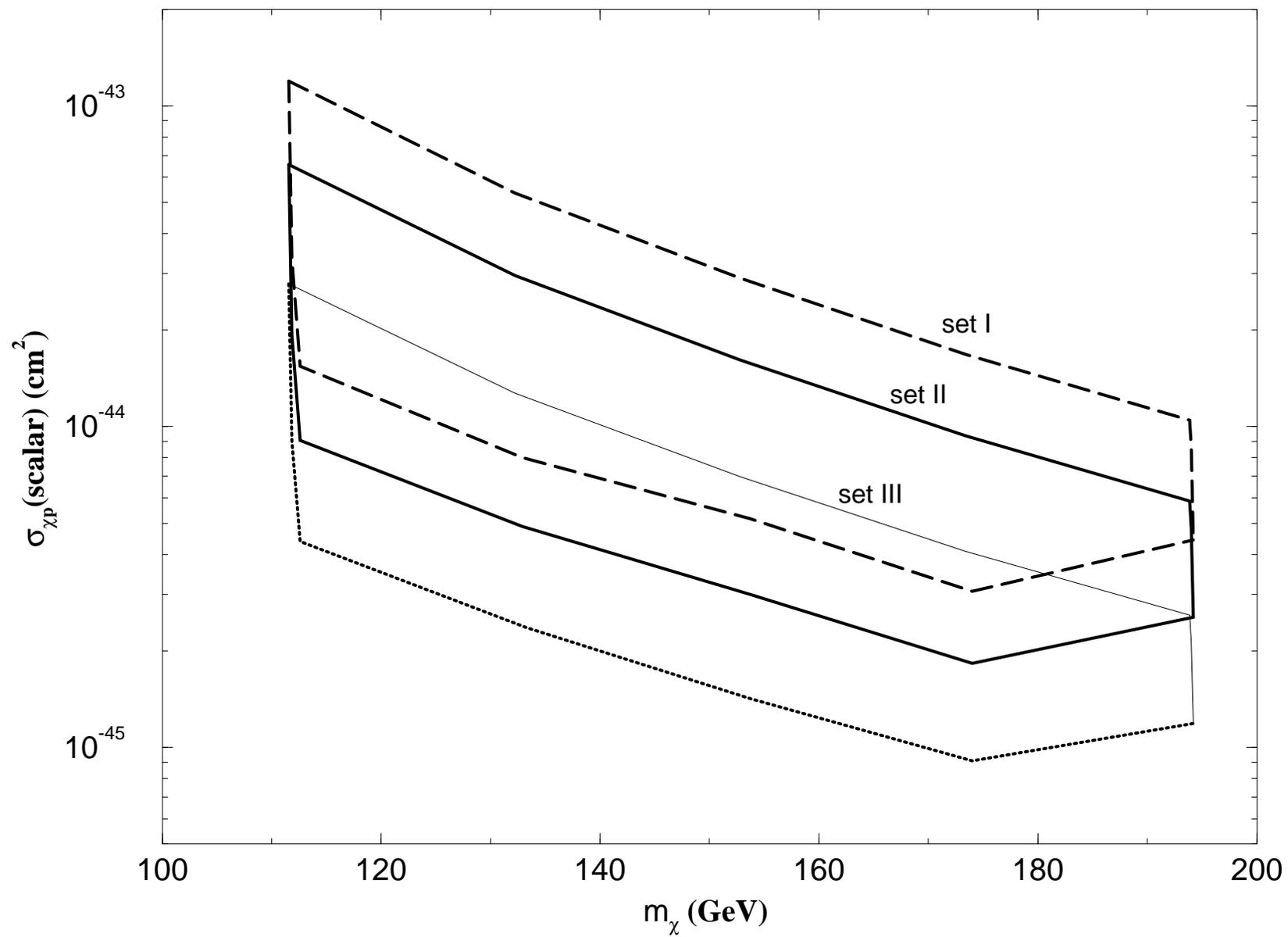
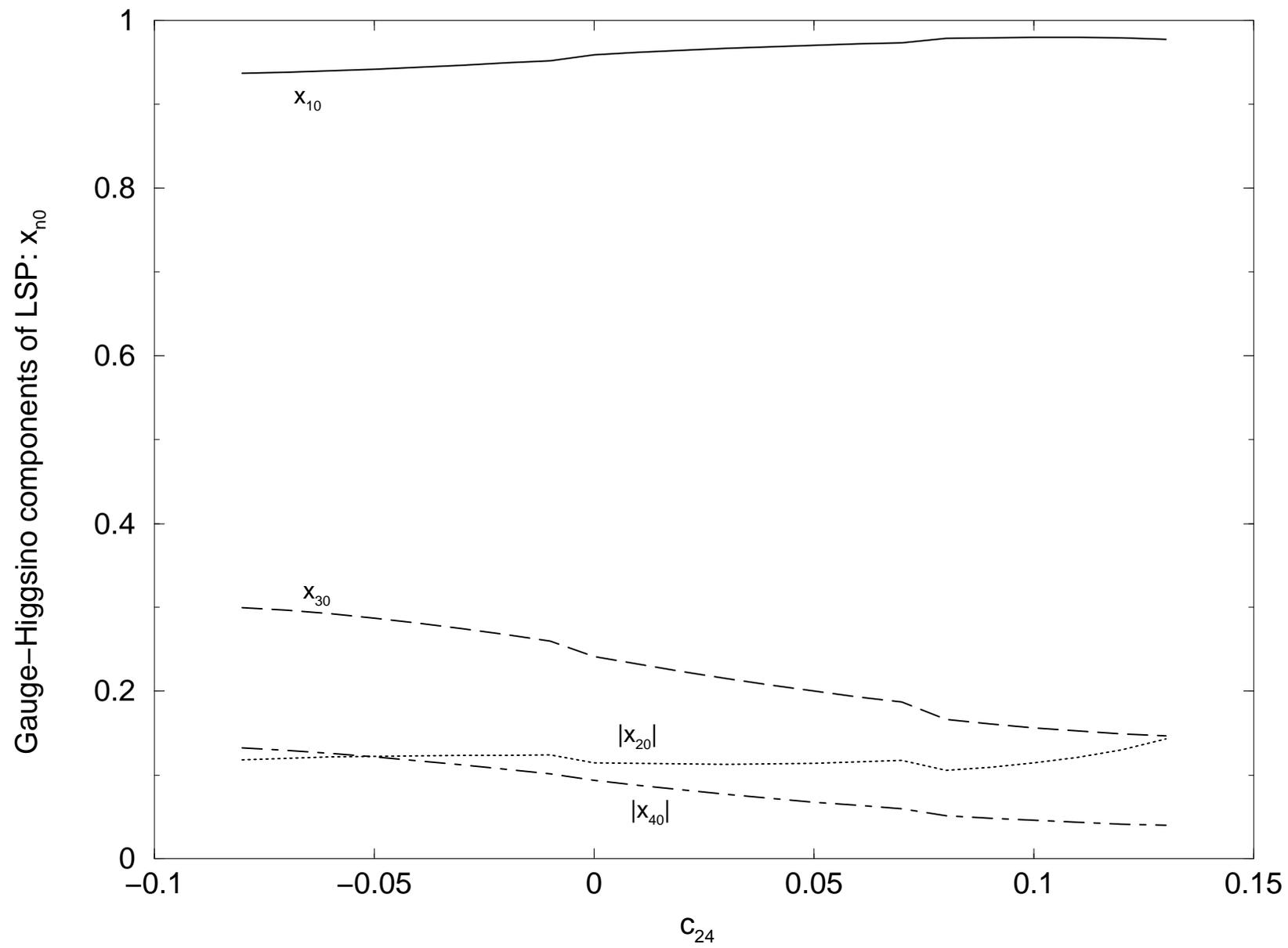
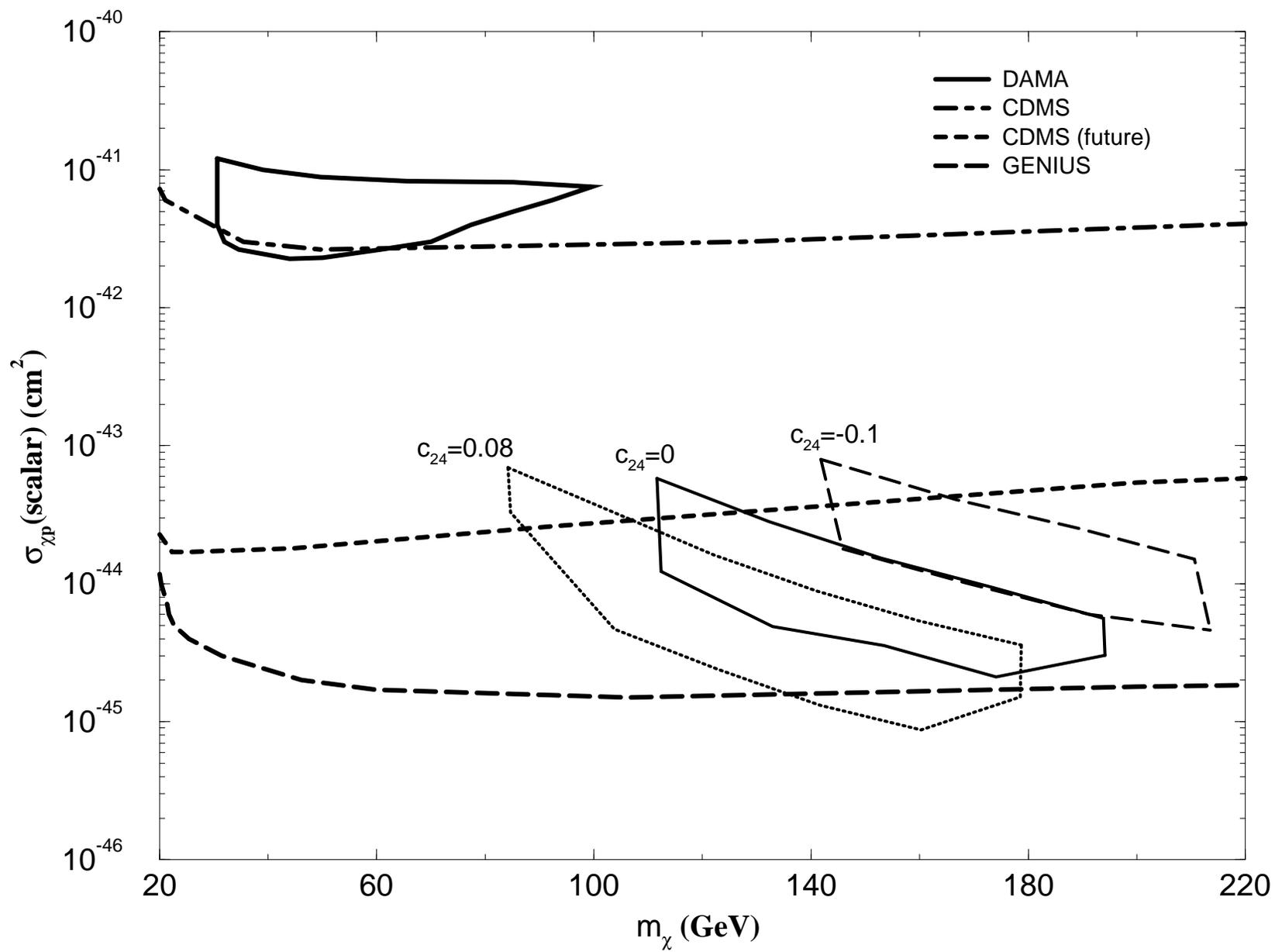
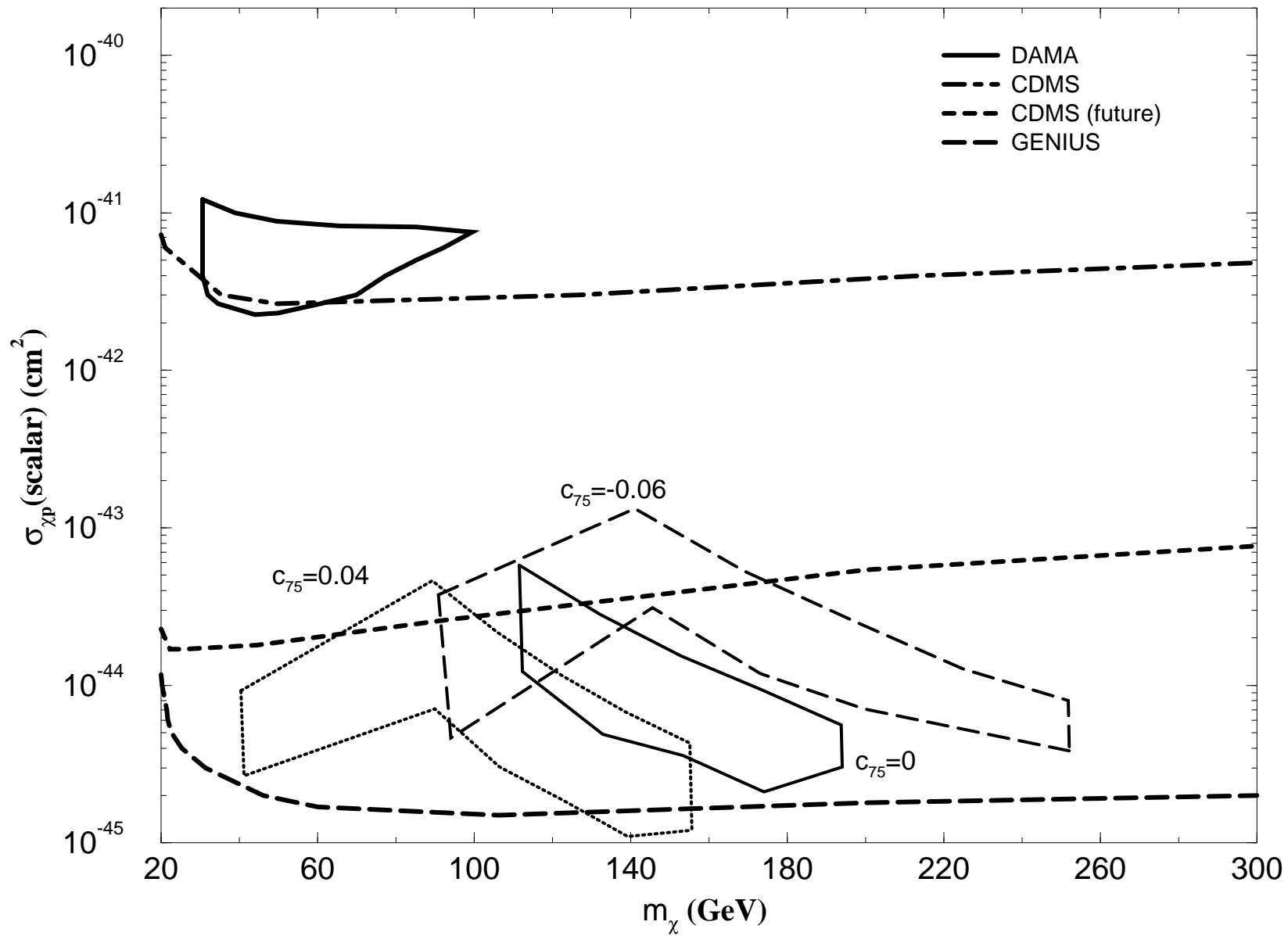
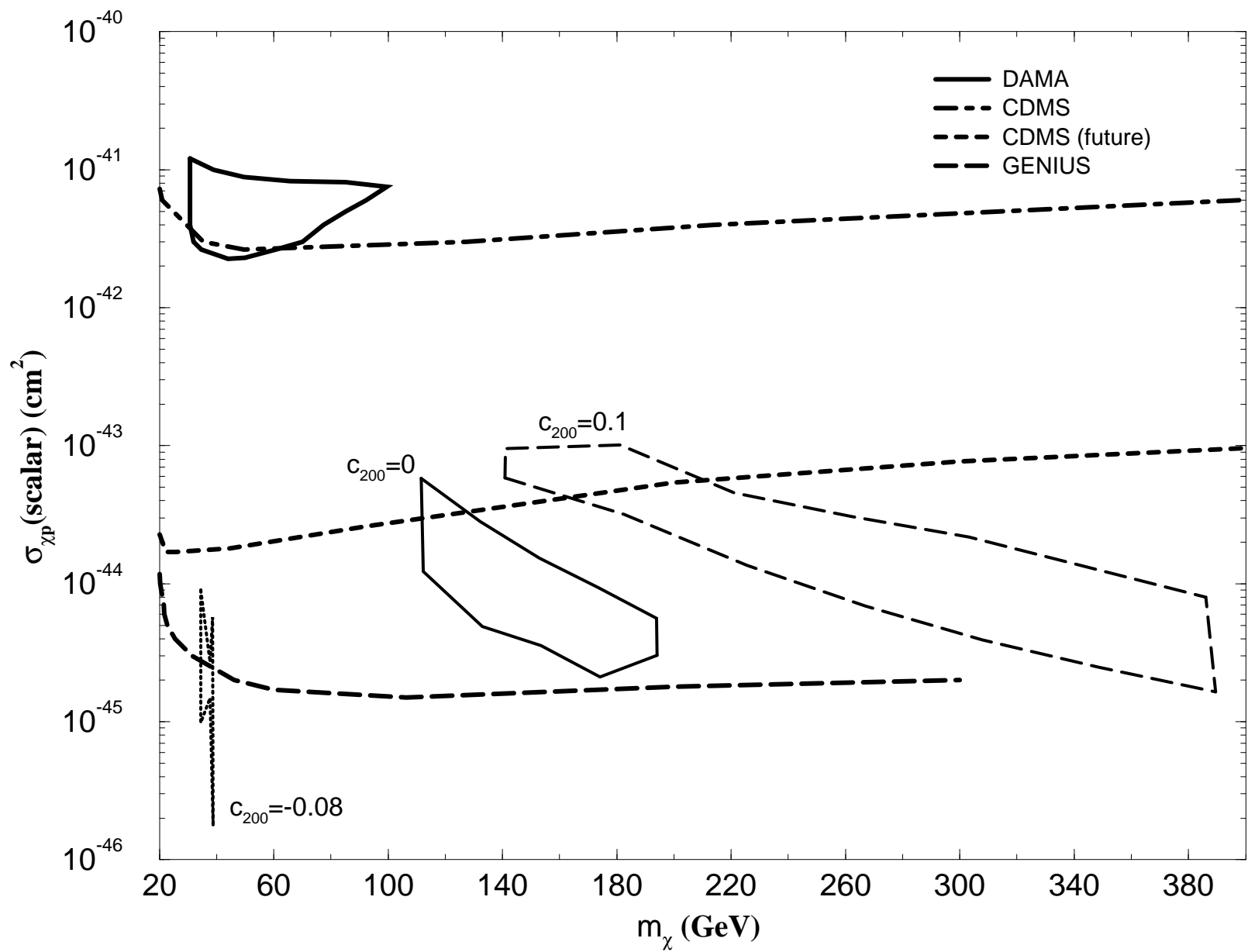
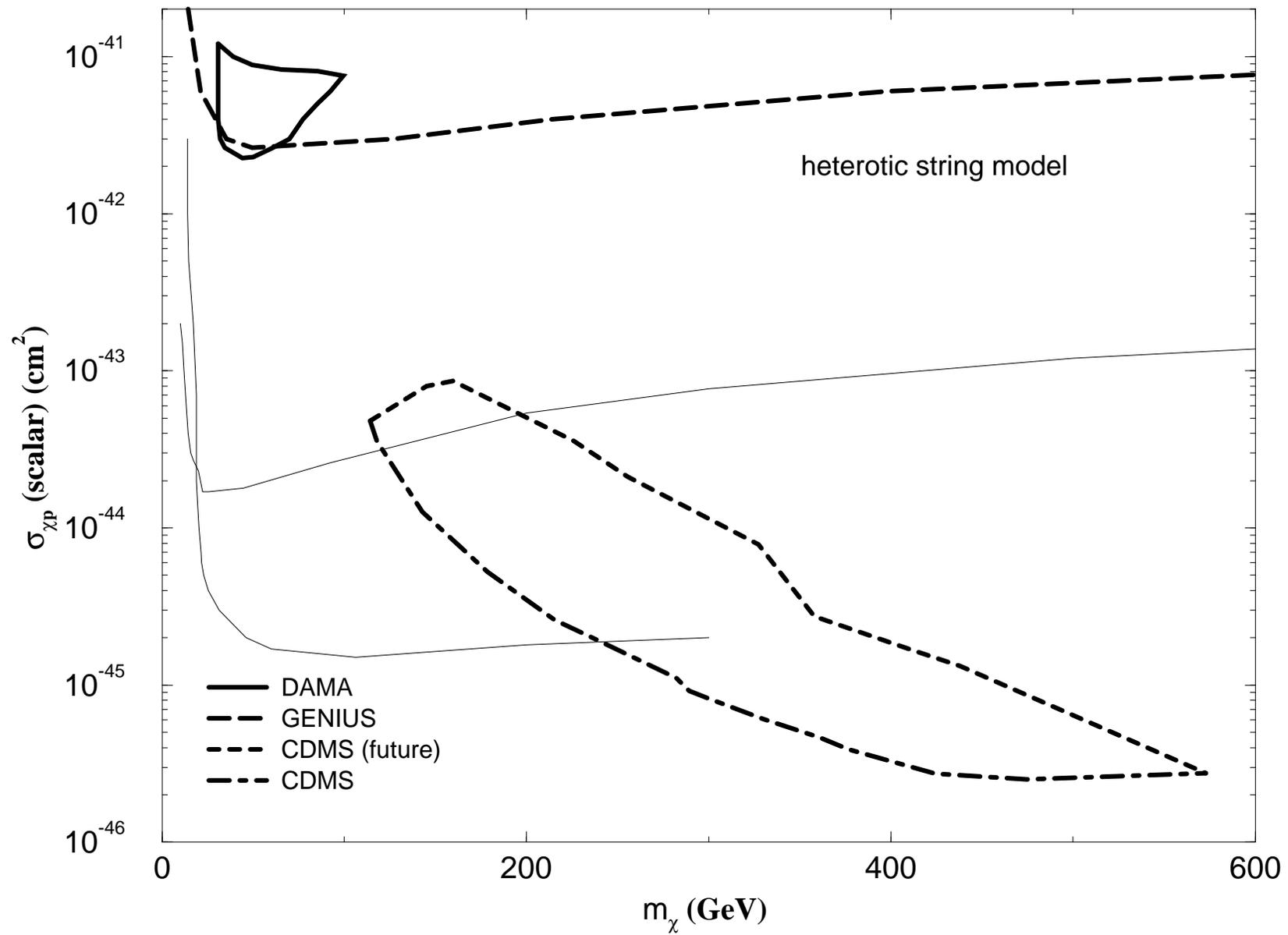
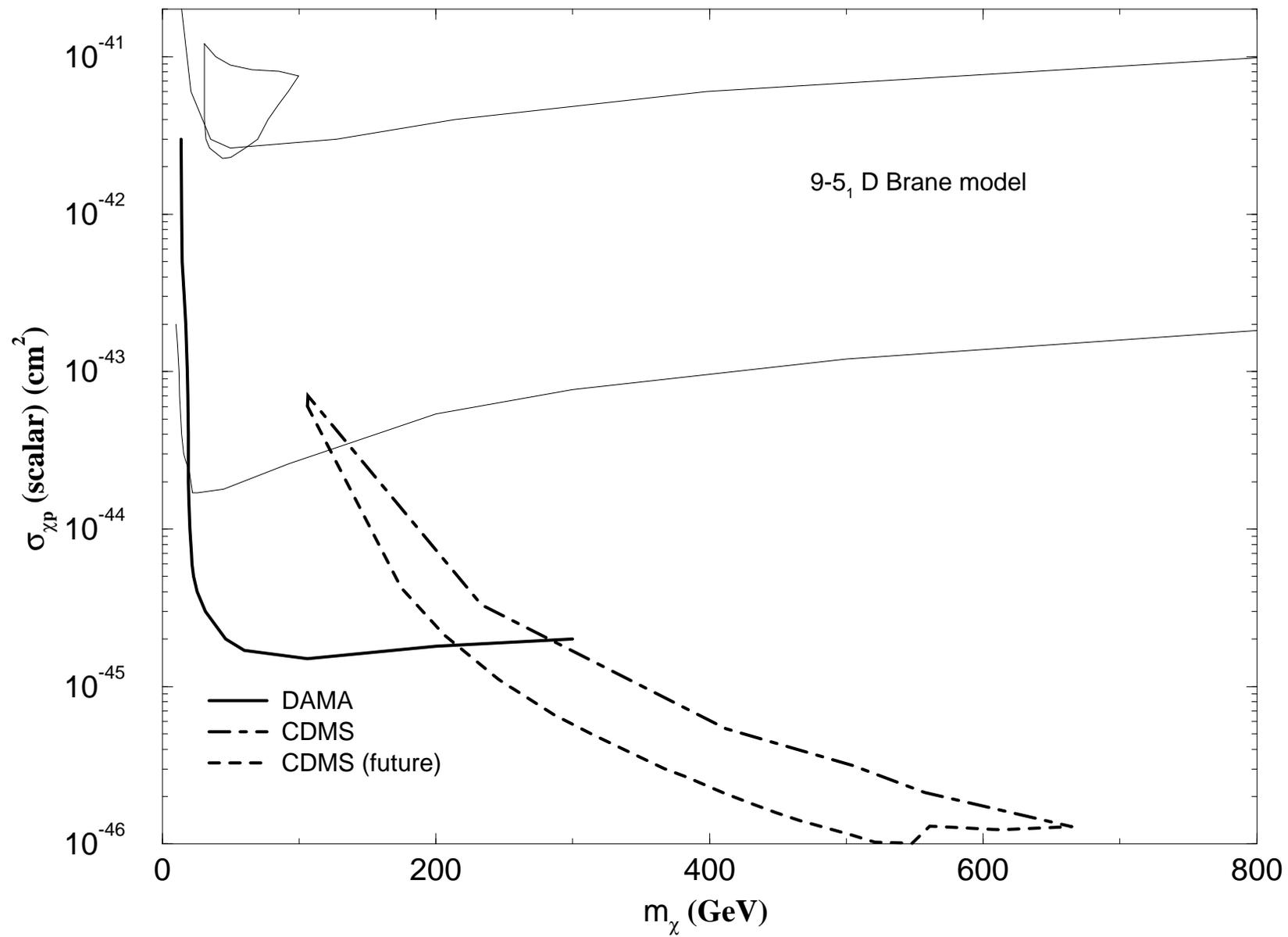
\end{document}